\documentclass[11pt]{article}
\usepackage{graphicx}
\usepackage{subfigure}

\begin{document}

\newcommand{\half}{\frac{1}{2}}
\newcommand{\inv}{^{-1}}

\title{De Sitter Thermodynamics from Diamonds's Temperature}
\author{Yu Tian \\{\it Institute of Theoretical Physics, Chinese Academy of Sciences}\\
{\it P. O. Box 2735, Beijing 100080}\\
{\tt ytian@itp.ac.cn}}

\maketitle

\begin{abstract}
The thermal time hypothesis proposed by Rovelli \cite{Rovelli}
regards the physical basis for the flow of time as thermodynamical
and provides a definition of the temperature for some special
cases. We verify this hypothesis in the case of de Sitter
spacetime by relating the uniformly accelerated observer in de
Sitter spacetime to the diamond in Minkowski spacetime. Then, as
an application of it, we investigate the thermal effect for the
uniformly accelerated observer with a finite lifetime in dS
spacetime, which generalizes the corresponding result for the case
of Minkowski spacetime \cite{MR}.

Furthermore, noticing that a uniformly accelerated dS observer
with a finite lifetime corresponds to a Rindler observer with a
finite lifetime in the embedding Minkowski spacetime, we show that
the global-embedding-Minkowski-spacetime (GEMS) picture of
spacetime thermodynamics is valid in this case. This is a rather
nontrivial and unexpected generalization of the GEMS picture, as
well as a further verification of both the thermal time hypothesis
and the GEMS picture.
\end{abstract}

\newpage

\section{Introduction}

The thermodynamics of spacetimes has drawn much interest after the
discovery of Hawking radiation \cite{Hawking} and Unruh effect
\cite{Unruh}. The thermodynamics of de Sitter (dS) spacetime
\cite{GH} is especially amazing and leads to a lot of puzzles
\cite{puzzles}, which are made more pressing by recent
cosmological observations showing that our universe is probably
asymptotically dS \cite{cosmo}.

Recently, Martinetti and Rovelli reviewed the Unruh effect from
the viewpoint of the so-called ``thermal time hypothesis"
\cite{Rovelli} and applied this hypothesis to study the Unruh
effect for an observer with a finite lifetime, which they call the
``diamonds's temperature" \cite{MR}. In this paper, we explore the
possibility of extending the application of this hypothesis to dS
spacetime. First, we find that it gives the correct (constant)
temperature for a uniformly accelerated observer in dS spacetime,
as the evidence of its validity in this case. Then, we go on to
apply it to a uniformly accelerated observer with a finite
lifetime in dS spacetime. Similar to the result in \cite{MR}, we
explicitly obtain a time-dependent temperature for this kind of
observers.

Unexpectedly, we find an elegant relation between our result and
the so-called global-embedding-Minkowski-spacetime (GEMS) picture
of spacetime thermodynamics \cite{DL,GEMS}. First, the temperature
for the uniformly accelerated dS observer is consistent with the
known result \cite{DL} obtained from the GEMS picture. Next,
noticing that a uniformly accelerated dS observer with a finite
lifetime corresponds to a Rindler observer with a finite lifetime
in the embedding Minkowski spacetime, we show that the GEMS
picture is valid in this case. In other words, the thermal time
hypothesis and the GEMS picture are compatible with each other, at
least in this case. Although both their physical meanings are not
very clear, this compatibility seems to justify both of them to
some extent.

\section{Thermodynamics of an eternal observer in dS spacetime from diamonds's temperature}
\label{sec:eternal}

\begin{figure}[htbp]
\centering
\includegraphics{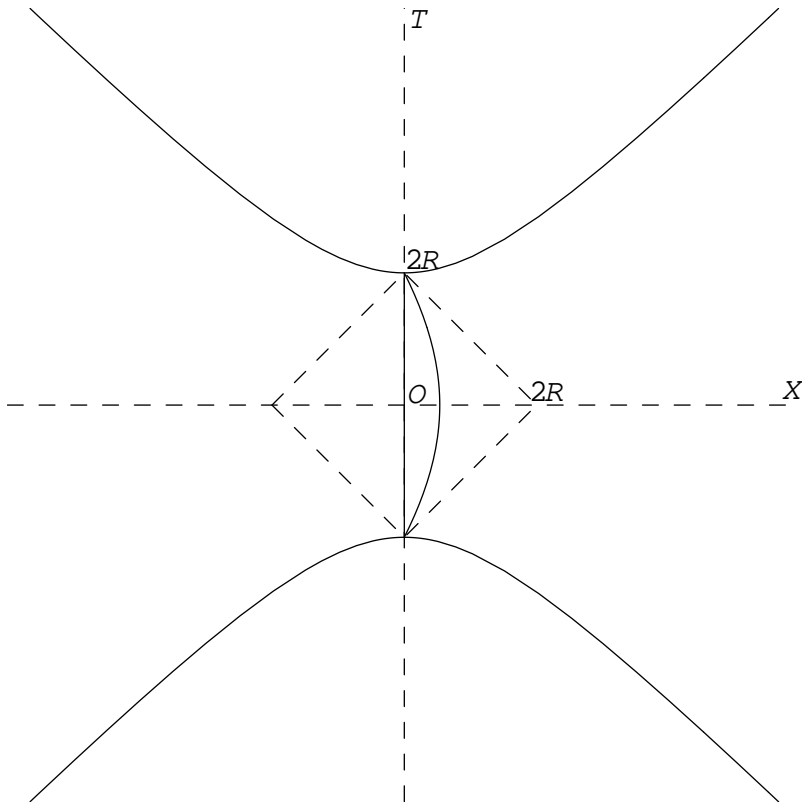}
\caption{A two-dimensional sketch map of the conformally flat
coordinates on the dS spacetime, where we have taken the notation
$T\equiv x^0$ and $X\equiv x^1$. All the points on the plane
except those on the hyperbola with equation
$c(x)=1-\frac{T^2-X^2}{4R^2}=0,$ which is actually the conformal
boundary of the dS spacetime, are points on the dS spacetime. The
diamond embraced by the dashed lines is the region covered by the
static coordinates. The solid line segment is the world line of
the inertial observer, while the solid segment of a hyperbola is
the world line of the observer staying at $r=R/2$. When $r\to R$,
the world line (\ref{world line}) of the uniformly accelerated
observer tends to the boundary of the diamond, $T^2=(X-2R)^2$.}
\label{CFdS}
\end{figure}

\begin{figure}[htbp]
\centering\subfigure[Penrose diagram of the dS spacetime, with
identification $AB=EF$, and with $ACE$ and $BDF$ its conformal
boundary. It is conformally compactified by identifying $AC=DF$
and $BD=CE$. The small diamond with vertices $C$ and $D$ is the
region covered by the static coordinates.] {\label{Penrose-dS}
\begin{minipage}{.8\textwidth}
\centering\includegraphics[scale=.8]{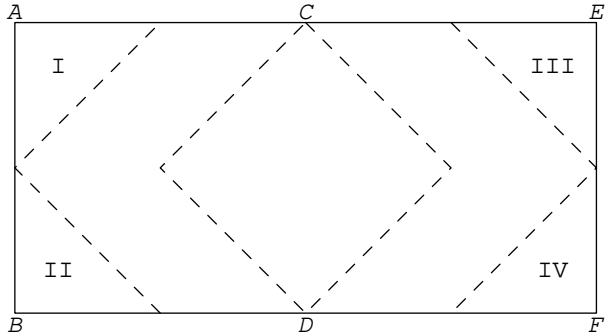}
\end{minipage}}\vspace{.4cm}
\subfigure[Penrose diagram of the Minkowski spacetime, which can
be compared with Fig.\ref{CFdS}, with the horizontal dashed lines
here corresponding to the conformal boundary of the dS spacetime
(the hyperbola) there. It is conformally compactified by
identifying $KL=MN$ and $KM=LN$. What the stereographic projection
does is just to remove the triangles (numbered ``I" to ``IV") in
Fig.\ref{Penrose-dS} to the corresponding positions in this
figure, respectively.] {\begin{minipage}{.8\textwidth}
\centering\includegraphics[scale=.8]{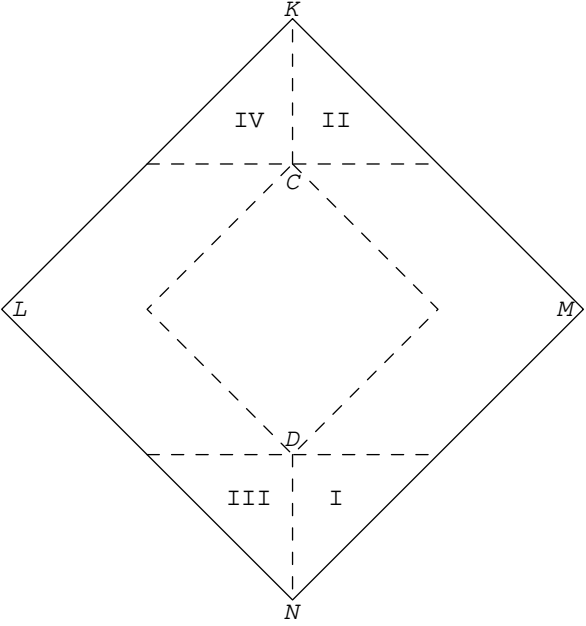}
\end{minipage}}\caption{An
illustration of the stereographic projection in terms of Penrose
diagrams.}\label{Penrose}
\end{figure}

The key observation is that the region covered by the static
coordinates, with metric
\begin{equation}
ds^2=(1-r^2/R^2)dt^2-\frac{1}{1-r^2/R^2}dr^2-r^2 d\Omega^2,
\end{equation}
on dS spacetime can be mapped to a diamond (which we call the dS
diamond) on Minkowski spacetime via a conformal transformation
which maps the dS spacetime to Minkowski spacetime in the sense of
conformal compactification.\footnote{Concerning the role played by
conformal transformations, \cite{MR} restricts itself to
conformally invariant quantum field theories. Since the spacetime
thermodynamics is mainly determined by the causal structure and
conformal transformations preserve the latter, in fact, one may
argue that the restriction can be removed.} From the 5-dimensional
viewpoint that regards the dS spacetime as a pseudo-sphere
\begin{eqnarray}
&&\eta_{AB}\xi^A\xi^B=-R^2, \\
&&ds^2=\eta_{AB}d\xi^A d\xi^B
\end{eqnarray}
with $\eta_{AB}=\mathrm{diag}(1,-1,-1,-1,-1)$, this conformal
transformation is simply realized by a (pseudo-)stereographic
projection,
\begin{equation}
x^\mu=\frac{2R\xi^\mu}{R+\xi^4}, \quad \mu=0,1,\cdots,3,
\end{equation}
from the point $P=(0,0,0,0,-R)$ to the hyperplane $\xi^4=R$ with
Minkowski signature, where $x^\mu$ stand for the coordinates
$\xi^\mu$ on this hyperplane. At the same time, $x^\mu$ can be
taken as coordinates on the dS spacetime, known as the conformally
flat coordinates. This coordinate system can cover almost the
entire dS spacetime, with metric
\begin{equation}\label{ds}
ds^2=\frac{\eta_{\mu\nu}dx^\mu dx^\nu}{c^{2}(x)}, \quad c(x)\equiv
1-\frac{\eta_{\mu\nu}x^\mu x^\nu}{4R^2},
\end{equation}
except the light cone at $P$. For a sketch map of the conformally
flat coordinates, see Fig.\ref{CFdS}. In terms of Penrose
diagrams, one can simply illustrate the relation between the
conformal compactificaion of the dS spacetime and that of the
Minkowski spacetime induced by the stereographic projection (see
Fig.\ref{Penrose}).

As is well known, the observer with world line $r=\mathrm{const}$
in the static coordinates is a uniformly accelerated observer in
the dS spacetime, with constant acceleration
\begin{equation}
a=\frac{r}{R (R^{2}-r^{2})^{1/2}}.
\end{equation}
Especially, the observer staying at $r=0$ is of acceleration
$a=0$, \emph{i.e.}, an inertial observer, and the acceleration
diverges when $r\to R$. In terms of the conformally flat
coordinates, the equation of the world line $r=\mathrm{const}$ is
\begin{equation}\label{world line}
T^2=X^2-\frac{4R^2}{r}X+4R^2,
\end{equation}
where we have restricted ourselves to the two-dimensional case for
simplicity. Taking $r$ as a free parameter, the above equation is
the set of all pseudo-circles\footnote{By the term pseudo-circle,
in fact, we mean a hyperbola with equation
$$(T-T_0)^2-(X-X_0)^2=W,$$ where $T_0$, $X_0$ and $W$ are
arbitrary real constants.} that passing the points $(0,-2R)$ and
$(0,2R)$. See Fig.\ref{CFdS} for the world lines of uniformly
accelerated observers in the conformally flat coordinates.

Now it is easy to obtain the action of the modular group (see
eq.(45) of \cite{MR}) on the dS diamond:
\begin{eqnarray}
T(\rho)&=&\frac{2R\sinh\rho}{\cosh\rho+R (R^2-r^2)^{-1/2}}, \label{T of rho}\\
X(\rho)&=&\frac{2R r
(R^2-r^2)^{-1/2}}{\cosh\rho+R(R^2-r^2)^{-1/2}}.
\end{eqnarray}
One can check that this modular flow is along the world line
(\ref{world line}), which is the precondition of a well-defined
temperature from the thermal time hypothesis. On the other hand,
we have from eq.(\ref{ds})
\begin{equation}
ds^2=(dT^2-dX^2)\Big(1-\frac{T^2-X^2}{4R^2}\Big)^{-2},
\end{equation}
which leads to
\begin{equation}\label{ds dT}
\frac{ds}{dT}=\frac{2R (R^2-r^2)^{1/2}[2R^2+(4R^4-4R^2 r^2+r^2
T^2)^{1/2}]}{(4R^2-T^2)(4R^4-4R^2 r^2+r^2 T^2)^{1/2}}.
\end{equation}
In order to apply the thermal time hypothesis \cite{MR}, we need
to obtain $d\rho/dT$, in terms of $T$, from eq.(\ref{T of
rho}).\footnote{It seems elementary to fulfill this task, but one
will encounter considerable difficulty if not appealing to some
algebraic tricks.} After some simplification, the final result is
\begin{equation}
\frac{d\rho}{dT}=\frac{2R [2R^2+(4R^4-4R^2 r^2+r^2
T^2)^{1/2}]}{(4R^2-T^2)(4R^4-4R^2 r^2+r^2 T^2)^{1/2}}.
\end{equation}
The thermal time hypothesis then gives
\begin{equation}\label{dS beta}
\beta=2\pi\frac{ds}{d\rho}=2\pi (R^2-r^2)^{1/2},
\end{equation}
which is independent of $T$ (constant for the observer) and in
exact agreement with the known result \cite{DL}.

\section{Thermodynamics of an observer with finite lifetime in dS spacetime from diamonds's temperature}

The above agreement, which at least shows some evidence of the
validity of the thermal time hypothesis in the dS spacetime,
stimulates us to go further along this direction, i.e., to
consider a uniformly accelerated observer with finite lifetime in
the dS spacetime, in order to generalize the result in \cite{MR}.
It can be shown that the static observer with lifetime $-\tau\le
t\le\tau$ in the static coordinates is associated with a reduced
diamond with half height
\begin{equation}
2M=\frac{2R\sinh(\tau/R)}{\cosh(\tau/R)+R (R^2-r^2)^{-1/2}},
\end{equation}
which is smaller than $2R$ and tends to $2R$ when $\tau\to\infty$.
 Now the modular flow corresponding to the reduced diamond is
\begin{equation}
T(\rho)=\frac{2M\sinh\rho}{\cosh\rho+\displaystyle\frac{R
(R^2-r^2)^{-1/2}\cosh(\tau/R)+1}{\cosh(\tau/R)+R
(R^2-r^2)^{-1/2}}},
\end{equation}
instead of eq.(\ref{T of rho}). Thus we can obtain $d\rho/dT$, by
making the replacement\footnote{It implies
$$r\to\frac{r\sinh^2(\tau/R)}{[\cosh(\tau/R)+(R^2-r^2)^{1/2}/R][\cosh(\tau/R)+R
(R^2-r^2)^{-1/2}]}.$$}
\begin{equation}
R\to M, \qquad R (R^2-r^2)^{-1/2}\to\frac{R
(R^2-r^2)^{-1/2}\cosh(\tau/R)+1}{\cosh(\tau/R)+R
(R^2-r^2)^{-1/2}},
\end{equation}
as
\begin{equation}
\frac{d\rho}{dT}=\frac{2R\{2R^2
[\cosh\frac{\tau}{R}+\frac{(R^2-r^2)^{1/2}}{R}]+F
[\cosh\frac{\tau}{R}+\frac{R}{(R^2-r^2)^{1/2}}]\}\sinh\frac{\tau}{R}}{\{4R^2\sinh^2(\tau/R)-T^2
[\cosh(\tau/R)+R (R^2-r^2)^{-1/2}]^2\}F},
\end{equation}
where we have defined
\begin{equation}\label{F}
F\equiv (4R^4-4R^2 r^2+r^2 T^2)^{1/2}.
\end{equation}
Then from eq.(\ref{ds dT}) and the thermal time hypothesis we have
\begin{equation}\label{beta}
\beta=\frac{2\pi
(R^2-r^2)^{1/2}(2R^2+F)\{4R^2\sinh^2\frac{\tau}{R}-T^2
[\cosh\frac{\tau}{R}+\frac{R}{(R^2-r^2)^{1/2}}]^2\}}{\{2R^2
[\cosh\frac{\tau}{R}+\frac{(R^2-r^2)^{1/2}}{R}]+F
[\cosh\frac{\tau}{R}+\frac{R}{(R^2-r^2)^{1/2}}]\}(4R^2-T^2)\sinh\frac{\tau}{R}}.
\end{equation}
This formula looks complicated, but we will see in the next
section that it can be thoroughly simplified in terms of the
static time $t$, with an elegant relation to the so-called GEMS
picture. In fact, the temperature has similar time dependence as
that in \cite{MR}, i.e., diverges at the two ends of the lifetime
($T=\pm 2M$) and has a minimum at $T=0$ with
\begin{equation}
\beta=2\pi (R^2-r^2)^{1/2}\tanh\frac{\tau}{2R}.
\end{equation}

As a primary check of the complicated expression (\ref{beta}), we
take its limit of $R\to\infty$ with $r$ fixed, and then get
\begin{equation}
\beta=\pi\frac{\tau^2-T^2}{\tau},
\end{equation}
which is actually the same as the $a\to 0$ limit of eq.(54) in
\cite{MR}. Its $a\ne 0$ case is not so straightforward, since in
this case $r$ tends to infinity as $R$ does, and then $T$ does not
tend to the proper time. First we have
\begin{equation}\label{s t}
s=(1-r^2/R^2)^{1/2}t=(1+a^2 R^2)^{-1/2}t
\end{equation}
with $s$ the proper time, and so
\begin{eqnarray}
T&=&\frac{2R\sinh(t/R)}{\cosh(t/R)+R
(R^2-r^2)^{-1/2}} \label{T t}\\
&\to &\frac{2\sinh(a s)}{a}.
\end{eqnarray}
Thus follows
\begin{equation}
F\to [4R^2 a^{-2}+4R^2 a^{-2}\sinh^2(a s)]^{1/2}=2R a^{-1}\cosh(a
s).
\end{equation}
Substituting the above equations into eq.(\ref{beta}) and taking
the $R\to\infty$ limit, we have finally
\begin{equation}\label{flat}
\beta=2\pi\frac{\cosh(a\sigma)-\cosh(a s)}{a\sinh(a\sigma)}
\end{equation}
with the proper time range
\begin{equation}\label{range}
-\sigma\le s\le\sigma,
\end{equation}
which is actually the same as eq.(54) in \cite{MR}. That shows
eq.(\ref{beta}) has the correct flat limit.

Although the above discussion assumes the spacetime dimension to
be four, it does not depend on that, so it can be easily extended
to any spacetime dimensions.

\section{Embedding of the dS observers in Minkowski spacetime}

The global-embedding-Minkowski-spacetime (GEMS) picture of
spacetime thermodynamics has been extensively studied in the
literature \cite{DL,GEMS,CTGS,CT}, since it was first confirmed in
the case of dS spacetime \cite{BD,NPT}, due to the key observation
that an inertial observer in a $d$-dimensional dS spacetime
corresponds to a Rindler observer in the $(d+1)$-dimensional
embedding Minkowski spacetime. This picture matches the Hawking
temperature detected by a stationary observer in a curved
spacetime with the Unruh temperature detected by the corresponding
observer in the GEMS of that spacetime.

Since it is obvious that a uniformly accelerated observer with a
finite lifetime in the $d$-dimensional dS spacetime corresponds to
a Rindler observer with a finite lifetime in the
$(d+1)$-dimensional GEMS and both their thermal effects have been
known, it is natural and interesting to ask whether these two
thermal effects can match each other.

For a Rindler observer with the finite (proper) lifetime
(\ref{range}) in the 5-dimensional Minkowski spacetime, we have
from eq.(\ref{flat})
\begin{equation}\label{5flat}
\beta=2\pi\frac{\cosh(a_5\sigma)-\cosh(a_5
s)}{a_5\sinh(a_5\sigma)}.
\end{equation}
On the other hand, for the uniformly accelerated observer staying
at $r=\mathrm{const}$ in the static dS spacetime, we have for its
GEMS \cite{DL}
\begin{equation}
a_5=(a^2+R^{-2})^{1/2}=(R^2-r^2)^{-1/2}.
\end{equation}
Recalling eq.(\ref{s t}) then reduces eq.(\ref{5flat}) to
\begin{equation}\label{beta GEMS}
\beta=2\pi
(R^2-r^2)^{1/2}\frac{\cosh(\tau/R)-\cosh(t/R)}{\sinh(\tau/R)}
\end{equation}
with the static time range $-\tau\le t\le\tau$. This expression is
much simpler than eq.(\ref{beta}), but they are actually the same
when taking into account the relation (\ref{T t}) between the
conformally flat time $T$ and the static one $t$, which may be a
little surprising. We leave the detailed computation in the
appendix. Note that the identification of eq.(\ref{beta}) and
eq.(\ref{beta GEMS}) is nontrivial, since we have obtained them
from different approaches: the former from the stereographic
projection and the latter from the GEMS picture. (Of course, both
of them depend on the thermal time hypothesis.)

It is striking that the GEMS picture of spacetime thermodynamics
is valid for a uniformly accelerated observer with a finite
lifetime in the dS spacetime. The GEMS picture has only been
confirmed in many static cases for a long time, until \cite{CTGS}
points out that it can be valid for general stationary motions
with not only the temperatures but also the chemical potentials
matched. At the same time, there is strong evidence that the GEMS
picture fails for non-stationary motions \cite{CT}. Now the
uniformly accelerated dS observer with a finite lifetime serves as
an example that the GEMS picture can be generalized in another
way.

\section{Concluding remarks}

The thermodynamics of dS spacetime itself (i.e., not with respect
to a specific observer) can be investigated from various
viewpoints: the event horizon \cite{GH}, the GEMS of (eternal)
static observers \cite{DL}, and the thermal time hypothesis for
(eternal) static observers as in \S\ref{sec:eternal}. All these
viewpoints lead to the same result (\ref{dS beta}), which strongly
justifies the puzzling dS thermodynamics. At the same time, this
can be regarded as a verification of the thermal time hypothesis
in a non-flat spacetime, the significance of which should be
explored further.

Noticing that a uniformly accelerated dS observer with a finite
lifetime corresponds to a Rindler observer with a finite lifetime
in the GEMS, we have shown that the GEMS picture of spacetime
thermodynamics is valid in this case, which is a whole new result
of this paper. On one hand, this result can be regarded as a
further (and rather unexpected) verification of both the thermal
time hypothesis and the GEMS picture. On the other hand, it is
very interesting to study whether the GEMS picture can be
generalized any further.

\section*{Acknowledgments}

I would like to thank Prof. Y. Ling and Dr. H.-Z. Chen for helpful
discussions. Prof. T. Padmanabhan reminds me to consider the GEMS
picture further, whom I would also like to thank. This work is
partly supported by NSFC under Grant No. 10347148.

\appendix

\section{Simplification of eq.(\ref{beta}) in terms of the static time $t$}

First, substitution of the relation (\ref{T t}) between the
conformally flat time $T$ and the static one $t$ into
eq.(\ref{beta}) gives
\begin{equation}\label{beta t}
\beta = \frac{{2\pi (R^2 - r^2)^{1/2}(2R^2 +
F)\{\sinh^2\frac{\tau}{R}[\cosh\frac{t}{R} + \frac{R}{(R^2 -
r^2)^{1/2}}]^2 - \sinh^2\frac{t}{R}[\cosh\frac{\tau}{R} +
\frac{R}{(R^2 - r^2)^{1/2}}]^2\}}}{{\{2R^2 [\cosh\frac{\tau}{R} +
\frac{(R^2 - r^2)^{1/2}}{R}] + F[\cosh\frac{\tau}{R} +
\frac{R}{(R^2 - r^2)^{1/2}}]\}\{[\cosh\frac{t}{R} + \frac{R}{(R^2
- r^2)^{1/2}}]^2 - \sinh^2\frac{t}{R}\}\sinh\frac{\tau}{R}}}.
\end{equation}
Then, substituting eq.(\ref{T t}) into the definition (\ref{F}) of
$F$ in terms of $T$, we have
\begin{eqnarray}
F&=&\sqrt{4R^2 (R^2 - r^2) + 4R^2 r^2\frac{{\sinh^2(t/R)}}{{[\cosh(t/R) + R(R^2 - r^2)^{ - 1/2}]^2}}} \nonumber\\
&=& 2R\frac{{\sqrt{(R^2 - r^2)[\cosh(t/R) + R(R^2 - r^2)^{ - 1/2}]^2 + r^2\sinh^2(t/R)}}}{{\cosh(t/R) + R(R^2 - r^2)^{ - 1/2}}} \nonumber\\
&=& 2R\frac{{\sqrt{R^2\cosh^2(t/R) + 2R(R^2 - r^2)^{1/2}\cosh(t/R) + R^2 - r^2}}}{{\cosh(t/R) + R(R^2 - r^2)^{ - 1/2}}} \nonumber\\
&=& 2R^2\frac{{\cosh(t/R) + (R^2 - r^2)^{1/2}/R}}{{\cosh(t/R) +
R(R^2 - r^2)^{ - 1/2}}},
\end{eqnarray}
which leads to
\begin{equation}
2R^2 + F = 2R^2\frac{{[\cosh(t/R) + R(R^2 - r^2)^{ - 1/2}] +
[\cosh(t/R) + (R^2 - r^2)^{1/2}/R]}}{{\cosh(t/R) + R(R^2 - r^2)^{
- 1/2}}}
\end{equation}
and
\begin{eqnarray}
&& 2R^2 [\cosh(\tau /R) + (R^2 - r^2)^{1/2}/R] + F[\cosh(\tau/R) + R(R^2 - r^2)^{ - 1/2}] \nonumber\\
&=& 2R^2 \frac{{[\cosh\frac{\tau}{R} + \frac{(R^2 -
r^2)^{1/2}}{R}][\cosh\frac{t}{R} + \frac{R}{(R^2 - r^2)^{1/2}}] +
[\cosh\frac{t}{R} + \frac{(R^2 -
r^2)^{1/2}}{R}][\cosh\frac{\tau}{R} + \frac{R}{(R^2 -
r^2)^{1/2}}]}}{{\cosh(t/R) + R(R^2 - r^2)^{ - 1/2}}}. \nonumber
\end{eqnarray}
On the other hand, we have
\begin{eqnarray}
&&[\cosh(t/R) + R(R^2 - r^2)^{ - 1/2}]^2 - \sinh^2(t/R) \nonumber\\
&=& 1 + 2R (R^2 - r^2)^{ - 1/2}\cosh(t/R) + R^2/(R^2 - r^2) \nonumber\\
&=& R (R^2 - r^2)^{ - 1/2}\{[\cosh(t/R) + R (R^2 - r^2)^{ - 1/2}]
+ [\cosh(t/R) + (R^2 - r^2)^{1/2}/R]\} \nonumber
\end{eqnarray}
and
\begin{eqnarray}
&&\sinh^2(\tau/R)[\cosh(t/R) + R(R^2 - r^2)^{ - 1/2}]^2 - \sinh^2(t/R)[\cosh(\tau/R) + R(R^2 - r^2)^{ - 1/2}]^2 \nonumber\\
&=&[\cosh^2(\tau/R) - 1][\cosh(t/R) + R (R^2 - r^2)^{ - 1/2}]^2 - [\cosh^2(t/R) - 1][\cosh(\tau/R) + R (R^2 - r^2)^{ - 1/2}]^2 \nonumber\\
&=& 2\cosh ^2(\tau/R)\cosh(t/R)R (R^2 - r^2)^{ - 1/2} + \cosh^2(\tau/R)R^2/(R^2 - r^2) \nonumber\\
&& - \cosh^2(t/R) - 2\cosh(t/R)R (R^2 - r^2)^{ - 1/2} - R^2/(R^2 - r^2) \nonumber\\
&& - 2\cosh^2(t/R)\cosh(\tau/R)R (R^2 - r^2)^{ - 1/2} - \cosh^2(t/R)R^2/(R^2 - r^2) \nonumber\\
&& + \cosh^2(\tau/R) + 2\cosh(\tau/R)R (R^2 - r^2)^{ - 1/2} + R^2/(R^2 - r^2) \nonumber\\
&=& R (R^2 - r^2)^{ - 1/2}[2\cosh(\tau/R)\cosh(t/R) + \cosh(\tau/R)(R^2 - r^2)^{1/2}/R + \cosh(t/R)(R^2 - r^2)^{1/2}/R \nonumber\\
&& + \cosh(\tau/R)R (R^2 - r^2)^{ - 1/2} + \cosh(t/R)R (R^2 - r^2)^{ - 1/2} + 2][\cosh(\tau/R) - \cosh(t/R)] \nonumber\\
&=& R (R^2 - r^2)^{ - 1/2}\{[\cosh(\tau/R) + (R^2 -
r^2)^{1/2}/R][\cosh(t/R) + R (R^2 - r^2)^{ - 1/2}] \nonumber\\
&& + [\cosh(t/R) + (R^2 - r^2)^{1/2}/R][\cosh(\tau/R) + R (R^2 -
r^2 )^{ - 1/2}]\}[\cosh(\tau/R) - \cosh(t/R)]. \nonumber
\end{eqnarray}
Finally, substitution of the above four expressions into
eq.(\ref{beta t}) gives
\begin{equation}
\beta=2\pi
(R^2-r^2)^{1/2}\frac{\cosh(\tau/R)-\cosh(t/R)}{\sinh(\tau/R)},
\end{equation}
where the other complicated factors have all reduced out. This
thoroughly simplified formula is exactly the same as eq.(\ref{beta
GEMS}).

\end{document}